\documentclass[namedreferences,hyperref,optionalrh]{spr-sola}
\usepackage{graphicx}        
\usepackage[T1]{fontenc}
\usepackage{amsmath}
\usepackage{lmodern}
\usepackage{color}           

\chardef\us=`\_

\begin{document}

\begin{frontmatter}
\title{Exploring an X1.8 flare and associated quasi-periodic pulsations using SDO and Aditya-L1 observations, alongside NFFF extrapolation}

\author[addressref={aff1},]{\inits{P.}\fnm{Pawan}~\snm{Kumar}\orcid{0009-0000-3324-2987}}

\author[addressref=aff1,corref,email={sainisanjay35@gmail.com}]{\inits{S.}\fnm{Sanjay}~\snm{Kumar}\orcid{0000-0003-4578-6572}}
\author[addressref=aff1]{\inits{S.}\fnm{Sadashiv}\orcid{0009-0003-0045-4387}}

\author[addressref={aff3},]{\inits{S.R.}\fnm{Sachin Rajkapoor}~\snm{Kanaujiya}\orcid{0009-0009-3048-0109}}

\author[addressref=aff2]{\inits{Y.K.}\fnm{Yogesh Kumar}~\snm{Maurya}}

\address[id=aff1]{Department of Physics, Patna University, Patna 800005, India}

\address[id=aff3]{Department of Physics, National Institute of Technology Calicut, Kozhikode-673601, Kerala, India}

\address[id=aff2]{Indian Institute of Astrophysics, 2nd Block, Koramangala, Bangalore 560034, India.}

\runningauthor{Kumar et al.}
\runningtitle{\textit{Solar Physics} Exploring an X1.8 flare and associated quasi-periodic pulsations}

\begin{abstract}
The paper seeks to explore the initiation of an X1.8 class solar flare (SOL2024-10-09T0124) and the related quasi-periodic pulsations (QPPs), which are situated within NOAA active region 13848. To achieve this, we perform a multi-wavelength analysis of the event utilizing data from the Solar Dynamic Observatory (SDO) and Aditya-L1. Importantly, the observations indicate the emergence of a sigmoidal brightening at the onset of the flare, alongside the evolution of a complex morphology of flare brightenings and ribbons around the peak time of the flare. Furthermore, QPPs are detected during the impulsive phase of the flare, exhibiting a dominant period of approximately 4 minutes in the hard X-ray light curve. In order to comprehend the initiation of the flare, we construct the coronal magnetic field at a pre-flare stage by employing the non-force-free-field (NFFF) extrapolation model. The extrapolated magnetic field illustrates the existence of various significant magnetic features, including a magnetic flux rope, a three-dimensional (3D) null, sheared arcade, and quasi-separatrix layers (QSLs) within the flaring region. The recurrent reconnection occurring at these features is proposed to play a role in the initiation of the flare as well as the associated QPPs.
\end{abstract}


\end{frontmatter}

\section{Introduction}
\label{intro}
Solar flares are defined as extremely energetic events occurring within the solar atmosphere. These events signify the abrupt and explosive discharge of a substantial amount of magnetic energy from the solar corona \citep{2011LRSP....8....6S}. Occasionally, it is noted that during solar flares, a considerable volume of magnetized plasma is ejected into the heliosphere, which is termed coronal mass ejections (CMEs). Flares that are linked to the CMEs are referred to as eruptive flares \citep{2001ApJ...559..452Z}.

Eruptive flares are particularly often associated with magnetic flux ropes, which are structures composed of twisted magnetic field lines encircling a central axis
\citep{2021PhPl...28b4502N,2022FrASS...939061K}. Observing a flux rope directly is challenging due to their lower plasma density. Nevertheless, the eruption of prominences and filaments offers compelling evidence that twisted magnetic structures may be present in the corona before flares occur
\citep{2012A&A...540A.127K}. Importantly, the existence of coronal sigmoids, which are S-shaped or inverse S-shaped configurations of coronal loops typically seen in extreme-ultraviolet (EUV) and soft X-ray (SXR) wavebands
\citep{1996ApJ...464L.199R}, is often regarded as a potential candidate for the development of flux ropes
\citep{2009ApJ...700L..83G,2011A&A...526A...2G,2012ApJ...759..105S}.

Importantly, magnetic reconnection is recognized as the essential physical mechanism that is underlying the occurrence of solar flares. Magnetic reconnection involves a change in magnetic topology, along with the transformation of magnetic energy into mass flow, heat, and the acceleration of charged particles \citep{Kumar2016,2023AdSpR..71.1856P}. However, the physics of magnetic reconnection, particularly within the intricate three-dimensional (3D) magnetic configurations found in flaring regions, is not yet fully explored \citep{2011LRSP....8....6S,2023AdSpR..71.1856P}.

In a three-dimensional magnetic configuration, areas exhibiting significant gradients in the mapping of field lines are known as quasi-separatrix layers (QSLs). These QSLs are identified as potential locations for magnetic reconnection \citep{1999A&A...351..707T,2002JGRA..107.1164T,2006SoPh..238..347A}. Significantly, QSLs are frequently associated with high electric current densities in the surrounding area \citep{2009ApJ...700..559M,2009ApJ...704.1288W}. The magnetic reconnection occurring within QSLs is generally associated with the observable slipping motion of magnetic field lines, which results from the ongoing exchange of magnetic connectivity driven by reconnection \citep{1995JGR...10023443P,2003JGRA..108.1285P,2006SoPh..238..347A,2012A&A...543A.110A,2013A&A...555A..77J}. Such reconnections are believed to be vital in the initiation of flares and the development of flare ribbons \citep{1997A&A...325..305D,2006SoPh..238..347A,2013A&A...555A..77J,2023A&A...677A..43P,2025SoPh..300...22K}.

Furthermore, quasi-periodic pulsations (QPPs) represent intriguing phenomena associated with flare emissions. QPPs are typically defined by a series of repetitive, impulsive, and sequential pulsations that can be seen in time-dependent intensity curves during solar flares \citep{2009SSRv..149..119N}. Specifically, QPPs denote a flare time series comprising at least three consecutive pulsations \citep{2018SSRv..214...45M,2023A&A...671A..79C,2024ApJ...971...29I}. The term ``quasi-periodic'' is due to the varying instantaneous periods of observed oscillations in flare light curves, which are interpreted as ``quasi-periods'' \citep{2019PPCF...61a4024N}. Initially, \citet{1969ApJ...155L.117P} identified and recorded the QPPs in hard X-ray (HXR) and microwave wavebands. Following this, these phenomena were observed across the entire spectrum of electromagnetic wavelengths; including radio/microwave, white light, ultraviolet (UV), EUV, SXR, HXR, H-alpha, Lyman-alpha, and gamma ($\gamma$)- rays \citep{2001ApJ...562L.103A,2005ESASP.596E..46F,2006ApJ...644L.149O,2008SoPh..247...77L,2008MNRAS.388.1899S,2009SSRv..149..119N,2011A&A...533A..61G, 2011A&A...533A.116Y,2016ApJ...819L..24Y,2016ApJ...827L..30H,2019ApJ...875...33H,2020STP.....6a...3K,2021ApJ...910..123C,2024A&A...684A.215C,2025A&A...695A.136B,2026ApJ..1003...78K}. 

The periods of QPPs observed in different wavebands range from seconds to minutes \citep{2018SSRv..214...45M,2021SSRv..217...66Z}. Moreover, the identified periods appear to be strongly linked to the observed waveband, instrumental sensitivity and its temporal resolution \citep{2024ApJ...970...77L}. The QPPs with the shorter periods (seconds to tens of seconds) are attributed to the fast sausage-mode oscillations or plasmoid dynamics in the reconnecting current sheet \citep{2010ApJ...723.1651K,2024ApJ...970...77L}. In contrast, the longer periodicities ranging from several minutes to tens of minutes, are more frequently associated with a quasi-periodic or "load/unload" magnetic reconnection regime \citep{2009SSRv..149..119N,2019A&A...621A.106T,2024ApJ...973...74K,DongLi2025}. However, these associations are not necessarily unique, as QPPs with comparable periods may arise from different physical mechanisms. The debate continues regarding which generation mechanism should be accountable for initiating a specific QPP event, primarily due to the fact that current observational data lacks sufficient information to differentiate between the various mechanisms \citep{2023BAAS...55c.181I,2024ApJ...970...77L}. Previous studies on QPPs have proposed several physical mechanisms underlying the QPPs, such as recurrent and intermittent magnetic reconnections, the periodic occurrence of magnetic reconnections, and the interaction of magnetohydrodynamic (MHD) waves with the reconnection site \citep{2010PPCF...52l4009N,2016ApJ...822....7K,2021SSRv..217...66Z,2026NatAs..10...54A,2026NatAs..10...13L}. However, the origin of QPPs remains not fully understood and requires further investigations \citep{2016SoPh..291.3143V,2018SSRv..214...45M}.

In this study, we investigate the onset of an X1.8 class eruptive flare and the associated  QPPs, hosted by NOAA active region (AR) 13848 on 2024 October 9. For this purpose, we utilize the multi-wavelength observations and the HXR light curve. In addition, we supplement the observations with a non-force-free-field (NFFF) extrapolation \citep{2010JASTP..72..219H} to examine the key magnetic features present in the flaring region and their role in the flare onset. Importantly, the extrapolated magnetic configuration documents the presence of a 3D magnetic null, a flux rope, and a sheared arcade enveloping the null and the rope in the flare kernel. 

This paper is arranged in the following manner. Section \ref{obs} documents the key features depicted in multi-wavelength and HXR observations of the flare and the associated QPPs. Section \ref{nfff-exp} provides the overview of NFFF extrapolation model and the  extrapolation of AR 13848. Section \ref{results} elaborates the key magnetic configuration obtained by the NFFF extrapolation of the faring region and explores their relation with observations of flare. Finally we summarize the key findings of this paper in Section \ref{summ}.

\section{Observations}

\label{obs}
\subsection{Multi-wavelength observations of the flare from AIA/SDO}
\begin{figure}
    \centering
    \includegraphics[width=1.0\textwidth]{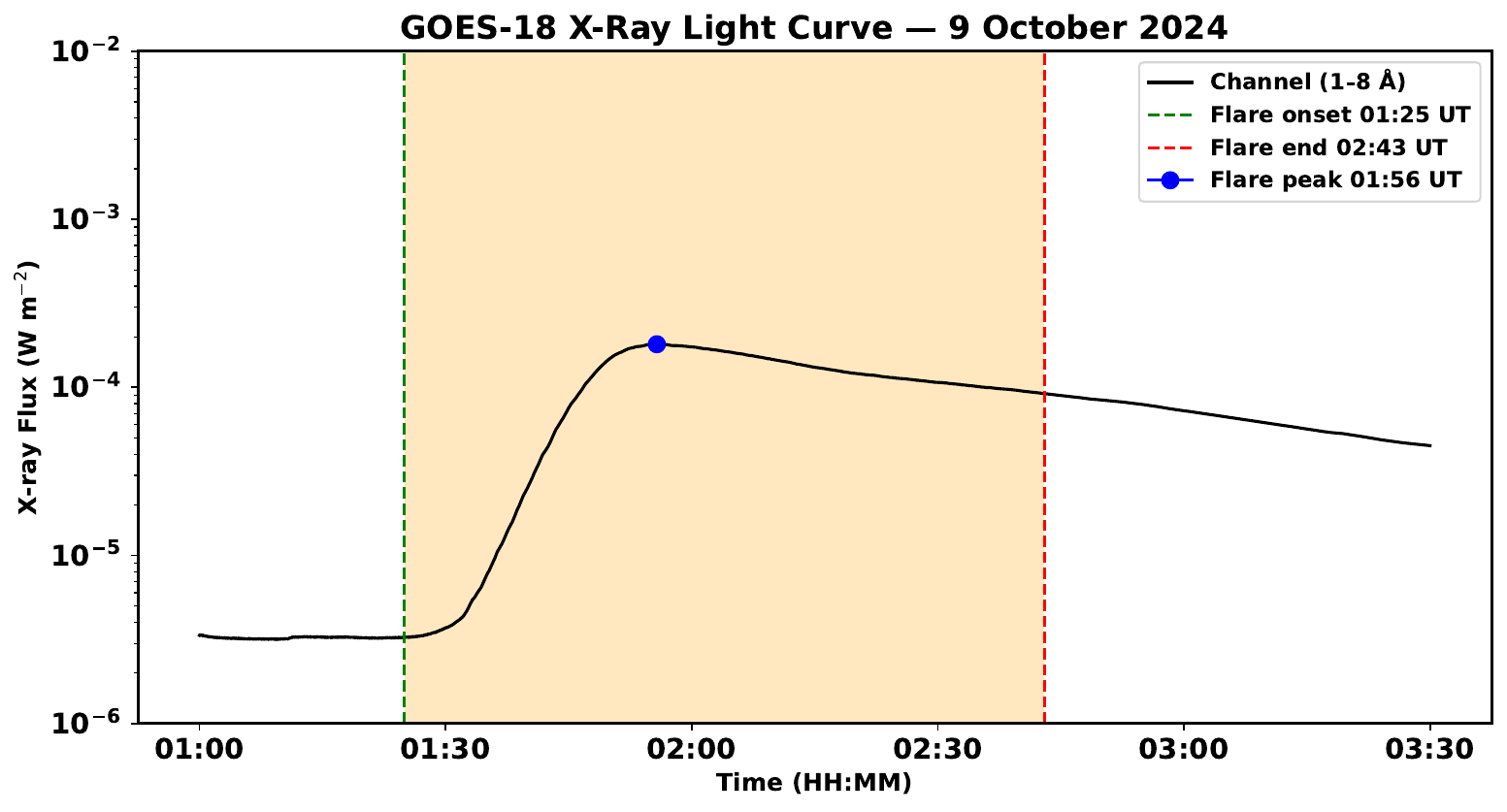}
    \caption{Figure shows the GOES-18 soft X-ray light curve in 1-8 {\AA} channel on 9 October 2024. The shaded region highlights the flaring interval bounded by two vertical dashed lines illustrated with green color (flare onset) and red color (flare end). The blue marker in the plot depicts the peak in soft X-ray flux.}
    \label{goes_flare}
\end{figure}

\begin{figure}
    \centering
    \includegraphics[width=1.0\textwidth]{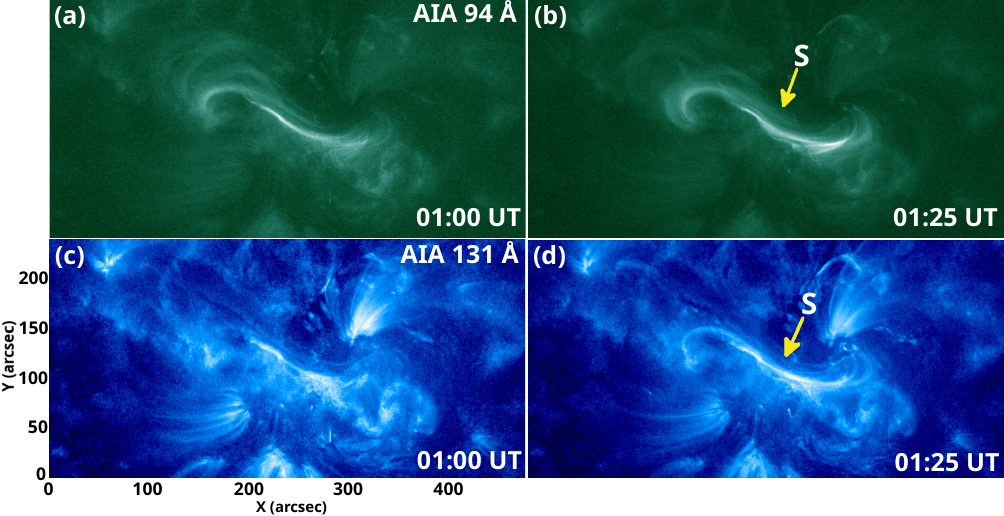}
    \caption{The flare evolution in AIA 94 {\AA} (panels (a) and (b)) and 131 {\AA} (panels (c) and (d)) channels. Panels (a) and (c) show the pre-flare stage at 01:00 UT, depicting the existence of S-shaped coronal loops before the flare onset in 94 {\AA} (panel (a)). Panels (b) and (d) document the appearance of sigmoidal brightening S (marked by yellow arrows) at the flare onset in both AIA 94 {\AA} and 131 {\AA}, respectively. }
    \label{94_131_obs}
\end{figure}

The X1.8 class flare occurred in NOAA AR 13848 on October 9, 2024. The flare started at 01:25 UT, peaked at 01:56 UT and ended at 02:43 UT in GOES observation plotted in Figure \ref{goes_flare}. The figure displays the soft X-ray light curve drawn from 01:00 UT to 03:30 UT, utilizing the X-ray flux data recorded by the GOES-18 X-ray sensor on October 9, 2024, with a temporal cadence of 1 second in 1-8 {\AA} channel. The shaded region encompasses the flare duration, whilst the dashed vertical lines in green and red color mark the flare onset ($\approx$ 01:25 UT) and end ($\approx$ 02:43 UT), respectively. The peak in the soft X-ray data is depicted by a blue marker at around 01:56 UT. The smooth rise to the peak followed by a gradual decay is consistent with the temporal evolution of soft X-ray emission during the selected flaring event. Further, to investigate the development of flare brightening and ribbons, we utilize the multi-wavelength observations from Atmospheric Imaging Assembly \citep[AIA,][]{2012SoPh..275...17L} instrument onboard the Solar Dynamic Observatory \citep[SDO,][]{2012SoPh..275....3P}. Figure \ref{94_131_obs} illustrates the evolution of flare from a pre-flare stage ($\approx$ 01:00 UT) to the flare onset ($\approx$ 01:25 UT) in AIA 94 {\AA} (panels (a) and (b)) and 131 {\AA} channels (panels (c) and (d)). The S-shaped bright structure in the region of interest in AIA 94 {\AA} (panel(a)) gets further brighten as time advances to the flare onset. This ultimately leads to the appearance of  the sigmoidal brightenings in both AIA 94 {\AA} and 131 {\AA} (marked with yellow arrows in panels (b) and (d)) at 01:25 UT.  

\begin{figure}
    \centering
    \includegraphics[width=1.0\textwidth]{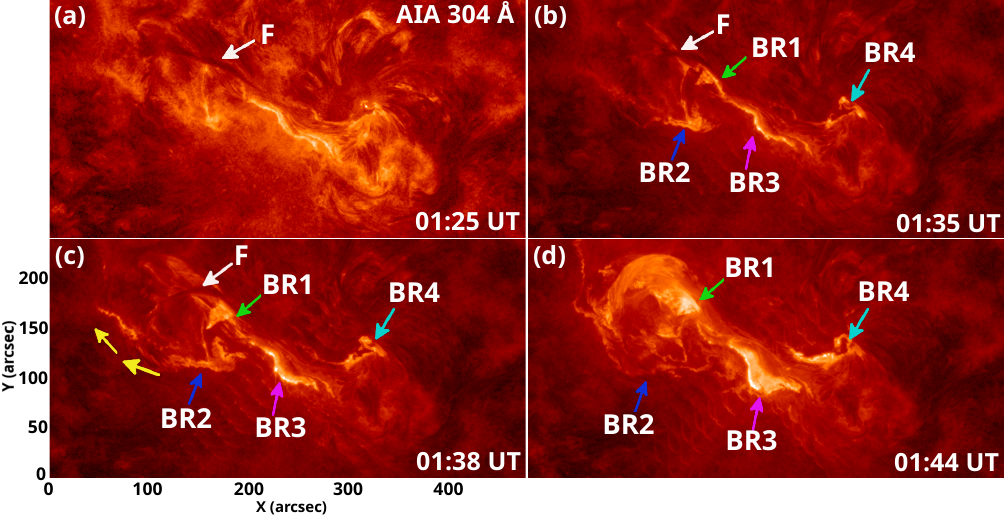}
    \caption{Time progression of flare brightenings in AIA 304 {\AA} starting from flare onset (01:25 UT). Panel (a) indicates the presence of a dark structure F (pointed by a white arrow). Panels (b) suggests the development of four flare brightenings: BR1 (green arrow), BR2 (blue arrow), BR3 (pink arrow) and BR4 (cyan arrow). Panel (c) shows the enhancement in flare brightenings and the expansion of BR2 in westward direction (yellow arrows). Panel (d) hints toward the disappearance of F. }
    \label{304_obs}
\end{figure}

We present the AIA 304 {\AA} observations in Figure \ref{304_obs}. Panel (a) shows the observation at the time of flare onset (01:25 UT). This illustrates a dark structure (F) indicated by a white arrow in the figure, signifying the existence of cool material suspended in the solar atmosphere during the pre-flare phase. Over time, four distinct flare brightenings emerge in the region. It is noteworthy that, initially, the brightenings BR1 and BR2 (denoted by green and blue arrows) are almost parallel (refer to panel (b)). Additionally, a serpentine brightening BR3 (indicated by a pink arrow) and a hook-shaped brightening BR4 (marked by a cyan arrow) are formed. Panel (c) illustrates that the aforementioned brightenings become increasingly bright over time, with BR2 expanding towards the west (highlighted by yellow arrows) and BR3 growing downward. Furthermore, as time progresses towards the flare peak, we observe the disappearance of F in panel (d). This suggests the eruption of F. The figure also indicates that BR2 diminishes, BR3 continues to grow downward, and BR4 expands westward over time.

\begin{figure}
    \centering
    \includegraphics[width=1.0\textwidth]{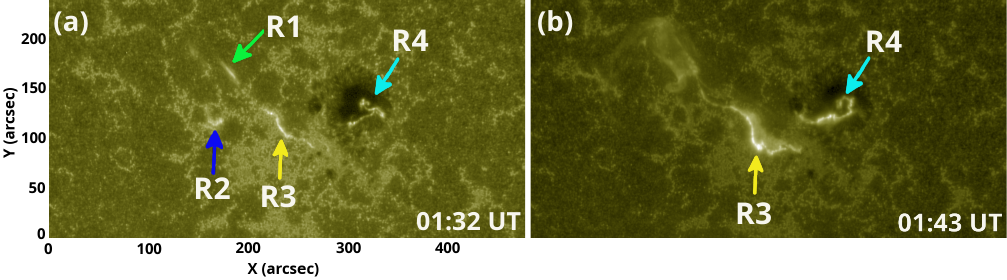}
    \caption{Flare observations in AIA 1600 {\AA} starting from 01:32 UT. Four flare ribbons R1 (green arrow), R2 (blue arrow), R3 (yellow arrow) and R4 (cyan arrow) are observed in the flare kernel in panel (a). R3 is characterized by a serpentine form, whereas R4 is shaped like a hook. In panel (b), R1 and R2 are no longer visible, while R3 extends downward and R4 stretches towards the west.}
    \label{1600_obs}
\end{figure}

In Figure \ref{1600_obs}, we present the observations of the flare captured in AIA 1600 {\AA}. Over time, four distinct flare ribbons are particularly noteworthy; R1 (indicated by a green arrow), R2 (marked by a blue arrow), R3 (marked by a yellow arrow), and R4 (indicated by a cyan arrow) at 01:32 UT in panel (a). Notably, R3 exhibits a serpentine structure, while R4 shows a hook-shaped morphology. At 01:43 UT (panel (b)), R3 appears to extend downward, and R4 extends towards the west as time progresses. The figure also illustrates the disappearance of R1 and R2 during this time interval. A comprehensive multi-wavelength analysis of the flare from various channels of AIA indicates that the four flare brightenings, BR1, BR2, BR3, and BR4, are nearly co-located with the flare ribbons R1, R2, R3, and R4 that developed in the flare kernel, respectively.

\subsection{QPPs observation from HEL1OS/Aditya-L1}

\begin{figure}
    \centering
    \includegraphics[width=1.0\textwidth]{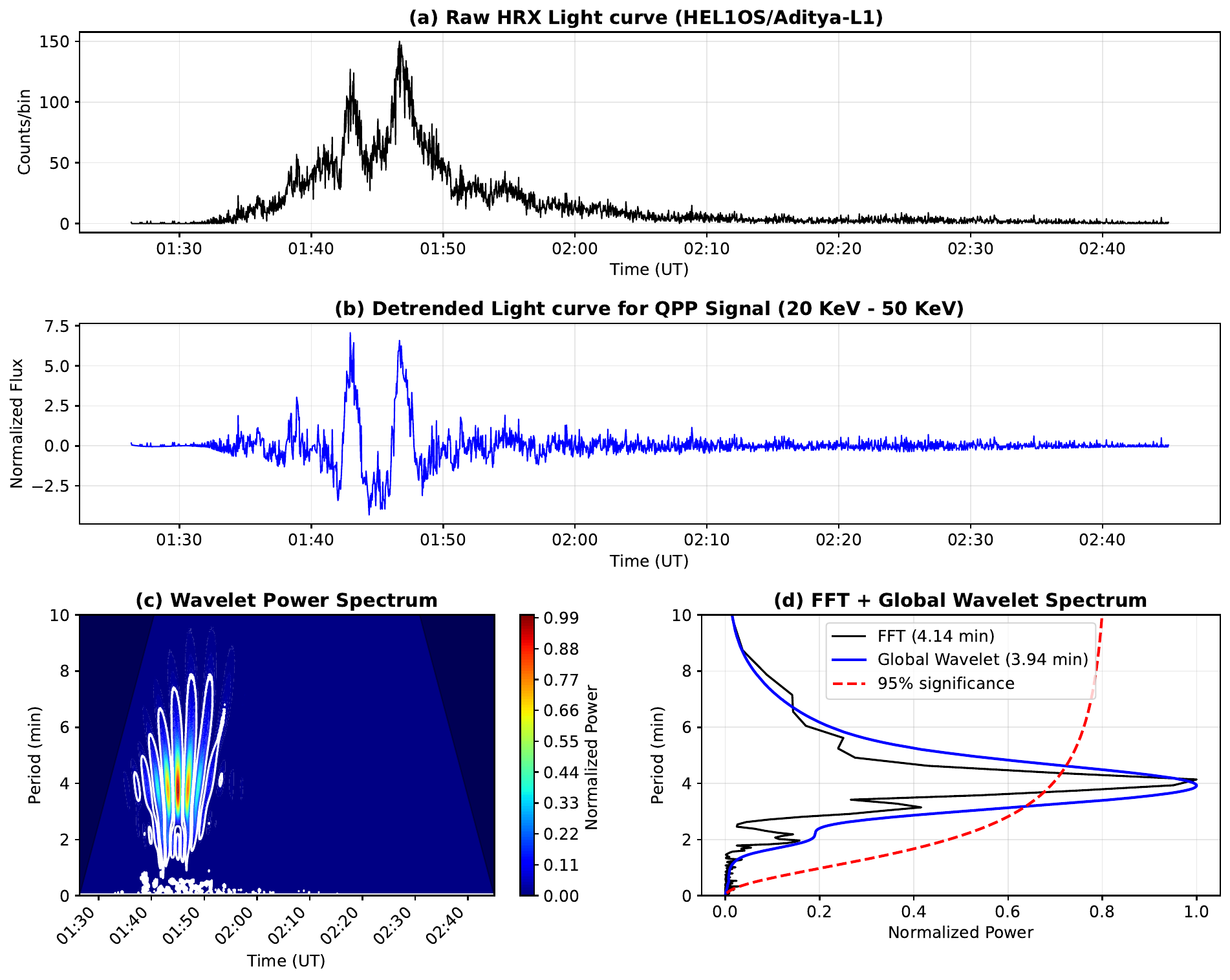}
    \caption{The HXR observation form the HEL1OS/Aditya-L1 in the 20–50 keV energy range. Panel (a) plots raw HXR light curve showing the temporal evolution of the flare emission. To obtain the QPP signal the normalized detrended intensity profile is plotted in panel (b). Panel (c) shows the Morlet wavelet power spectrum of the detrended signal. The white contours indicate the 95$\%$ confidence level estimated against the red-noise background depicting increased wavelet power is concentrated near a period of $\approx$ 4 minutes. Panel (d) compares the normalized Fourier power spectrum (black) and global wavelet spectrum (blue). The dashed red curve denotes the 95$\%$ significance level derived from the AR(1) red-noise model. Both spectral techniques reveal a dominant and statistically significant periodicity $\approx$ 4 minutes, confirming the presence of quasi-periodic pulsations in the HXR emission.}
    \label{QPP}
\end{figure}

To diagnose the QPPs associated with flare emission in HXR waveband within an energy range 20 keV to 50 keV, we have used the HXR light curve supplied by the High Energy L1 Orbiting X-ray Spectrometer \citep[HEL1OS,][]{2025SoPh..300..140N} onboard Aditya-L1 \citep{2025SoPh..300..128P}. Panel (a) of Figure \ref{QPP} illustrates the time binned (bin size = 2 second) raw light curve covering the full flare duration starting from 01:25 UT. Next to eliminate low-frequency fluctuations and to achieve stable background trends, we have applied a Savitzky–Golay smoothing filter \citep{savitzky1964smoothing} to the raw light curve with a window size of 251 seconds. This offers a detrended intensity profile that is appropriate for identifying the signature of QPPs and conducting the corresponding periodicity analysis, as shown in panel (b). The panel documents the presence of more than three peaks in the detrended light curve during the impulsive phase of the flare ($\approx$ 01:35 UT to $\approx$ 01:50 UT).

Next we have employed the Morlet wavelet transform following the method of \citet{1998BAMS...79...61T} to find the periodicity of QPPs in the detrended HXR light curve. Additionally, the uncertainty associated with the period derived from wavelet analysis is assessed based on the duration during which the global wavelet power spectrum exceeded the 95{\%} significance threshold. Figure \ref{QPP}(c) displays the wavelet power spectrum of the detrended 20–50 keV HXR light curve, indicating a notable increase in power at a period of $\approx$ 4 minutes. The white contours denote the 95$\%$ confidence level against the red-noise background, confirming that the observed oscillatory signal is statistically significant. Hence, the wavelet analysis reveals the signature of QPPs with a period of $\approx$ $4 \pm 0.7 $ minutes. To further confirm, another method is also used to determine the period of the QPPs. In this method, the Fourier analysis of the detrended light curve is performed using Fast Fourier Transform (FFT) periodogram. The uncertainty in the FFT period is determined by the full width at half maximum of the primary spectral peak which discloses a prominent peak corresponding to a period of $\approx$ $4 \pm 0.6 $ minutes. This indicates the presence of QPPs in the 20–50 keV HXR emission, consistent with the periodicity detected in the wavelet analysis. Figure \ref{QPP}(d) presents the normalized Fourier power spectrum (black curve) superimposed with the global wavelet spectrum (blue curve) derived from the detrended HXR light curve in the 20–50 keV energy band.
Both the spectra peak around the period $\approx$ 4 minutes --- suggesting the existence of a dominant periodicity $\approx$ 4 minutes. The close agreement between the two independent spectral techniques confirms the presence of the QPP signal. Also the dashed red curve represents the 95$\%$ confidence level estimated using a red-noise model. The dominant enhancement in the FFT and the global wavelet spectrum exceeds this significance threshold, indicating that the detected oscillation are not the red-noise fluctuations. The persistence of the spectral peak across both Fourier and wavelet analysis support the interpretation of a statistically significant QPPs during the impulsive phase of the flare.

\begin{figure}
    \centering
    \includegraphics[width=1.0\linewidth]{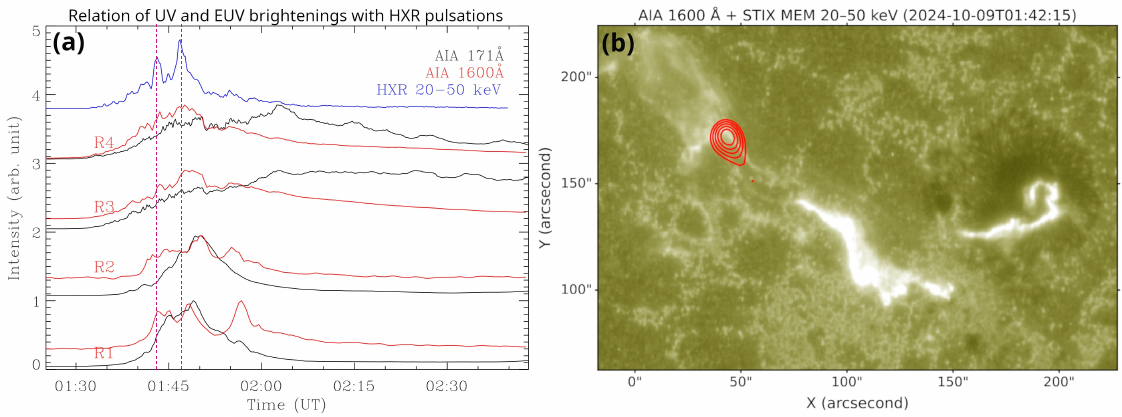}
    \caption{Figure illustrates the relation between the pulsations in HXR emission with the time series of EUV and UV emissions from the flaring regions. Panel (a) plots the temporal evolution of each flare brightening (R1, R2, R3 and R4) in AIA 171 {\AA} (EUV) and 1600 {\AA} (UV). The panel is further overlaid with the HXR emission (blue curve). Notably, the pulsations in HXR emission nearly coincide the pulsations in EUV and UV emissions from R1 (vertical dashed lines). Panel (b) shows the 20-50 keV HXR source contours with 60{\%} of maximum intensity overlaid on 1600 {\AA} image and co-spatial with the R1.  }
    \label{euv_uv_hxr_qpp}
\end{figure}

Furthermore, to explore the relation between the HXR pulsations and the flare brightenings/ribbons, in Figure \ref{euv_uv_hxr_qpp}(a), we plot the intensity profile of the locations of each flare brightening/ribbon observed in EUV and UV channels (highlighted by BR1, BR2, BR3 and BR4 in Figure \ref{304_obs} and R1, R2, R3, and R4 in Figure \ref{1600_obs}).  It is crucial to highlight that various EUV channels of AIA, with the exception of 171 {\AA}, experience significant intensity saturation during the progression of the chosen X class flare. Therefore, we select the AIA 171 {\AA} channel for the time series (plotted in black curves in panel (a)) analysis of the EUV flare brightenings. In addition to this, AIA 1600 {\AA} images are utilized to construct the temporal progression of the flare ribbons in UV channel (represented by the red curves in panel (a)).  Following \citet{Clarke_2021}, panel (a) is superimposed with the HXR light curve (blue curve) in the 20-50 keV energy range to facilitate a comparison between the EUV and UV intensity profiles and HXR emissions. It is significant to emphasize that the prominent peaks observed in the HXR light curve, are nearly coincident with the peaks in intensity profiles of R1 in both EUV and UV channels (see vertical red dashed lines in panel (a)).

To identify the location of the detected QPPs in the HXR emission within 20-50 keV energy band, we perform imaging analysis using the data from Spectrometer/Telescope for Imaging X-rays \citep[STIX,][] {2020A&A...642A..15K} onboard Solar Orbiter \citep{2020A&A...642A...1M}. We have utilized the $\mathtt{stixpy}$ package for the processing and visualization of the STIX data. Also, for the construction of HXR source, $\mathtt{MEM\_GE}$ algorithm \citep{2020ApJ...894...46M} incorporated in the $\mathtt{stixpy}$ package is employed. In particular, the HXR source is reconstructed within the 20-50 keV band by utilizing a 30 second integration window from 01:42:00 UT to 01:42:30 UT that coincides with the identified QPP signals. A suitable background with no heightened activity is chosen for the construction of the source region. Subsequently, the constructed 20-50 keV HXR source is reprojected on earth viewpoint using $\mathtt{SunPy}$ \citep{2020ApJ...890...68S} tools. The HRX source contours (in red color) are overlaid on AIA 1600 {\AA} image at 01:42 UT in panel (b) of Figure \ref{euv_uv_hxr_qpp}. Notably, the reprojected contours are manually co-aligned to the brightenings observed in 1600 {\AA} \citep{2024A&A...684A.215C,2026ApJ..1003...95L}. The contours represent the 60{\%} of maximum intensity of the constructed HXR source and are co-spatial with the flare ribbon R1 (see Figure \ref{1600_obs}). Relevantly, such HXRs are generally generated by non-thermal bremsstrahlung originated from the interaction of flare-accelerated electrons with the dense chromosphere \citep{1971SoPh...18..489B}. The near simultaneous peaks observed in EUV, UV emissions from the the ribbon R1 and the HXR emission (Figure \ref{euv_uv_hxr_qpp}(a)), along with the spatial alignment of the HXR contours with the ribbon R1 (Figure \ref{euv_uv_hxr_qpp}(b)), indicate that the R1 within the flaring kernel acts as a potential source region for the QPPs associated to the flaring event. In contrast, the emissions in the EUV and UV wavebands from other flare ribbons (including R2, R3, and R4) show a significantly lower correlation with the HXR emissions.

\section{Non-Force-Free-Field Extrapolation}
\label{nfff-exp}

In order to diagnose the key magnetic topologies in the flaring region underlying the X1.8 class flare onset, here we employ the NFFF extrapolation model to construct the coronal magnetic field configuration for AR 13848. This model was developed by \citet{2008ApJ...679..848H} and \citet{2010JASTP..72..219H}. The model operates on the principle of Minimum Dissipation Rate (MDR), which dictates that the plasma relaxes to a state that lowers the rate of energy dissipation. The relaxed state is described by the double-curl Beltrami equation and the magnetic field $\mathbf{B}$ is then expressed as \citep{2007SoPh..240...63B} 

\begin{equation}\label{double_curl_eq}
     \mathbf{B}=\mathbf{B}_{1}+\mathbf{B}_{2}+\mathbf{B}_{3} ; \nabla \times \mathbf{B}_{i}=\alpha_{i}\mathbf{B}_{i}
\end{equation}

\noindent{where $i=1,2$ and $3$, each $\mathbf{B}_{i}$ shows the linear-force-free-field associated with a corresponding twist parameter $\alpha_i$. In practical NFFF extrapolation, we choose $\alpha_2 = 0$ and $\alpha_1 \neq \alpha_3$, which makes $\mathbf{B}_{2}$ a potential field. For a comprehensive explanation and understanding of the NFFF extrapolation model, the readers are referred to \citet{2008SoPh..247...87H,2008ApJ...679..848H,2010JASTP..72..219H}.}

We use the photospheric vector magnetograms supplied by Helioseismic and Magnetic Imager \citep[HMI,][]{2012SoPh..275..229S} onboard SDO to perform the NFFF extrapolation. In particular we select the vector magnetic field data from the \textit{hmi.sharp\_cea\_720s} data series that have a pixel resolution $0.5''$ with a time cadence of 12 minutes. The magnetogram is first remapped onto the Lambert cylindrical equal area (CEA) projection, later it is converted into heliographic coordinates \citep{1990SoPh..126...21G} to derive the magnetic field within a Cartesian grid. The extrapolation is performed with aforementioned data series at 01:24 UT (1 minute prior to flare onset). To concentrate on the flaring region, the original magnetogram cutout is further cropped to $952 \times 476$ grid points along the $x$ and $y$-directions, respectively. Meanwhile $476$ grid points are utilized in the $z$-direction. This corresponds to a physical domain $\approx$ 342 Mm in $x$ direction and $\approx $171 Mm in both $y$ and $z$-directions. 

\section{Results}
\label{results}

\begin{figure}[t]
    \centering
    \includegraphics[width=1\textwidth]{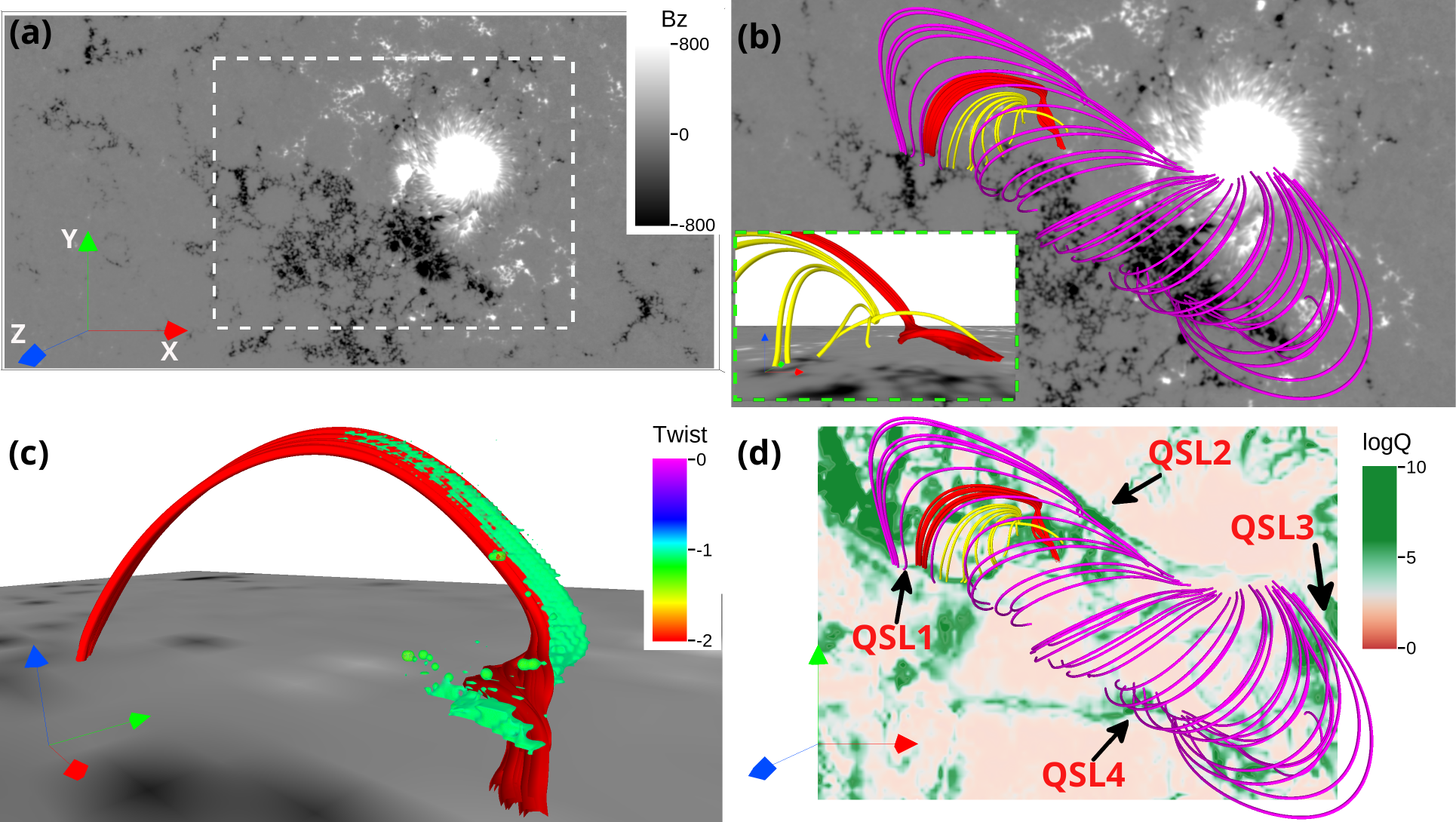}
    \caption{Panel (a) presents a line of sight vector magnetogram (Bz) at the lower boundary of AR 13848, with the white box highlighting the flaring area. Panel (b) depicts the extrapolated configuration of magnetic field lines within the flaring region, where the pink lines indicate a sheared arcade, the yellow lines correspond to a 3D null (as shown in the inset of the panel), and the red lines illustrate low-lying twisted magnetic structure. Panel (c) overlays the red field lines with the twist parameter, indicating that the red structure has a magnitude of the twist greater than 1. Panel (d) demonstrates the field lines superimposed with the squashing factor at the lower boundary, revealing that the foot-points of the field lines are anchored in regions of high Q. }
    \label{initial_exp}
\end{figure}

In Figure \ref{initial_exp}, we plot the extrapolated magnetic field line configuration at 01:24 UT representing the pre-flare stage. Panel (a) of the figure plots the line of sight component (Bz) of photospheric magnetic field at the bottom boundary ($z=0$) in the computational volume. Here, the flaring region in which flare brightenings are occurring is enclosed by the white dashed box. Panel (b) illustrates the magnetic field lines overlaid with Bz at the bottom boundary in the flaring region. The field lines plotted with pink represent the sheared arcade. The yellow field lines underneath the sheared arcade construct a 3D null with a dome shaped fan and anchored spine structure (inset of panel (b)). We have verified the presence of 3D null using the updated trilinear method of null detection \citep{2007PhPl...14h2107H,2020A&A...644A.150O, Maurya2024}. The red colored field lines located beneath the sheared arcade seem to construct a twisted magnetic structure similar to a magnetic flux rope. To confirm, in panel (c) these field lines are further overlaid with the twist parameter ($T_w$). In particular, $T_{w}$ is evaluated using the equation,
\begin{equation}
    T_{w}={\frac{1}{4 \pi} \int_l {\alpha} dl} 
\end{equation}
\noindent where {$\alpha$ = $\mathbf{\frac{J.B}{|B|^{\text{2}}}}$} and $dl$ is a line element \citep{2006JPhA...39.8321B}. Importantly, the twist parameter exceeding a magnitude of 1 is found to 
be co-located with the red field lines. This observation validates that the red magnetic structure signifies a flux rope situated beneath the sheared arcade \citep{2022FrASS...939061K}. Furthermore, it is important to note that the right leg of the flux rope is enveloped by the dome of the 3D null (see the inset of panel (b)).

In panel (d) the field lines are superimposed with the squashing factor \citep[Q,][]{2002JGRA..107.1164T} on a logarithmic scale at the bottom boundary which is a measure of field line mapping. We have calculated Q using the code by \citet{2016ApJ...818..148L}. It is important to note that the foot-points of the extrapolated field lines are anchored in the high $logQ$ regions, which direct towards the steep gradient in field line connectivity. Specifically, the areas with elevated $logQ$ values are recognized as QSLs and are regarded as potential locations for magnetic reconnection  \citep{1999A&A...351..707T,2002JGRA..107.1164T,2006SoPh..238..347A,2025SoPh..300...22K}. In this context, we can broadly identify four distinct QSLs, labeled as QSL1, QSL2, QSL3, and QSL4 in the figure.

\begin{figure}
    \centering
    \includegraphics[width=0.5\textwidth]{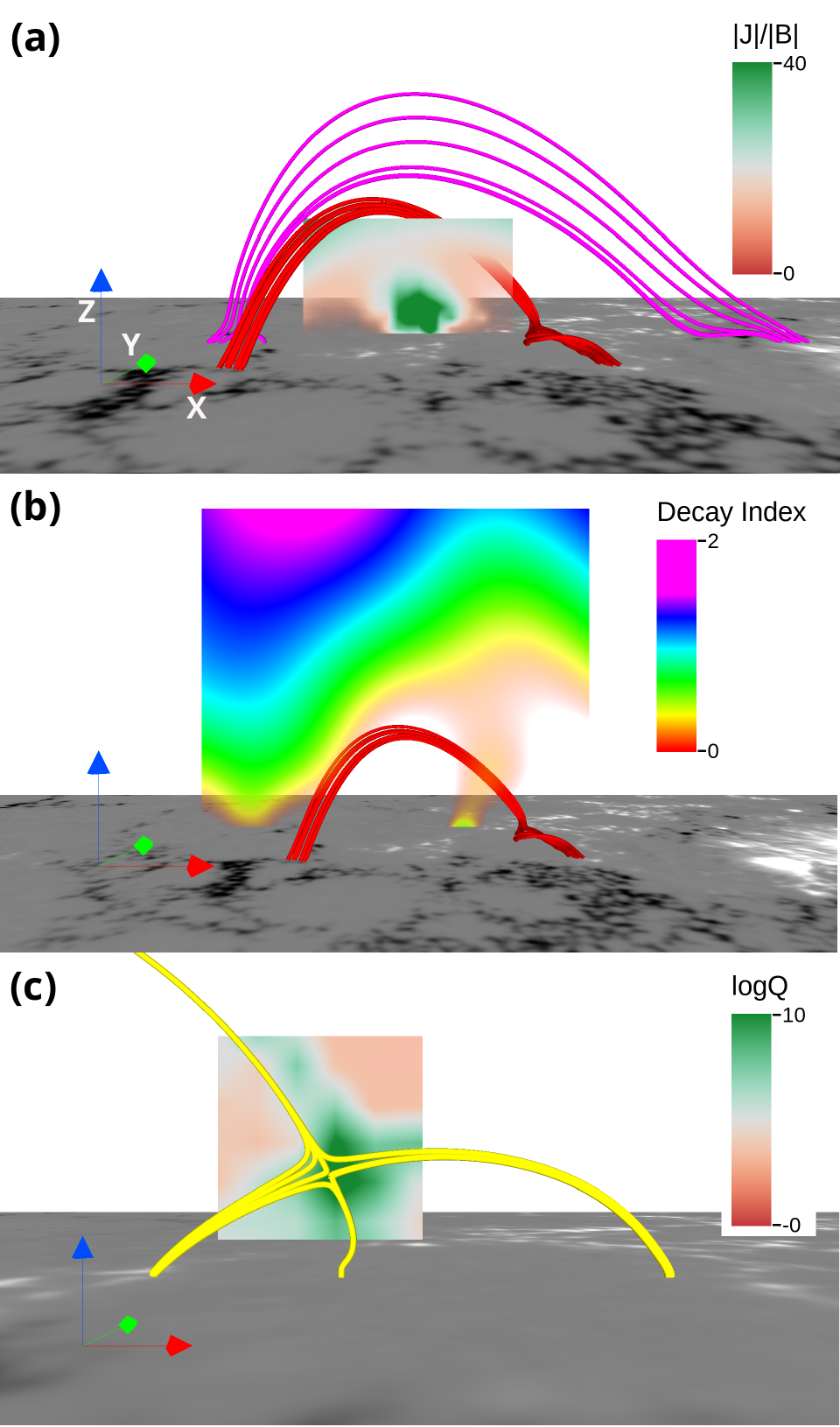}
    \caption{Panel (a) shows the projected field lines of the sheared arcade in a $y$-constant plane overlaid with $|\textbf{J}|/|\textbf{B}|$ and the flux rope. Notable is the high value of $|\textbf{J}|/|\textbf{B}|$ (beneath the flux rope), which indicates high magnetic field gradient in the sheared arcade. Panel (b) plots the flux rope overplotted with decay index in a $y$-constant plane --- depicting the torus unstable region over the flux rope. Panel (c) illustrates the $logQ$ superimposed with the projected field of the 3D null in a $y$-constant plane .  }
    \label{recn}
\end{figure}

To investigate the potential mechanism behind the initiation of the X1.8 flare, in panel (a) of Figure \ref{recn} we present the transverse field derived by setting $B_y=0$ in $\mathbf{B}$ for the pink field lines, which illustrate the projected field lines of the sheared arcade within a $y$-constant plane. Additionally, the panel is superimposed with $|\textbf{J}|/|\textbf{B}|$ in the $y$-constant plane, along with the field lines of the flux rope. It is noteworthy that there is a significantly elevated value of $|\textbf{J}|/|\textbf{B}|$ beneath both the sheared arcade and the flux rope. This indicates a high gradient in the magnetic field of the sheared arcade. One can envision a scenario where the field lines of the sheared arcade, anchored in opposite polarities, experience slipping motion within the QSLs (Figure \ref{initial_exp}(d)). This could promote an increase in the magnetic field gradient and potentially trigger reconnection within the sheared arcade \citep{2016ApJ...828...62J,2018NatCo...9..174I}. Furthermore, such reconnections are anticipated to occur repeatedly over time and may play a role in the ascent of the flux rope \citep{2016ApJ...830...80K,  Kumar2017Chaotic, 2023A&A...677A..43P}.

Figure \ref{recn}(b) illustrates the flux rope superimposed with the decay index (n) in a $y$-constant plane. The decay index measures the rate at which horizontal field is decaying with height and it is expressed as, 
\begin{equation}
    n=-{\frac{d \ln B_{ex}}{d \ln h}}
\end{equation}
\noindent{where $B_{ex}$ represents the horizontal component of the external magnetic field that overlays the flux rope at height $h$. A critical value of $n\approx 1.5$ represents the torus unstable region \citep{2006PhRvL..96y5002K}. Significantly, the torus unstable region is situated above the flux rope. This leads to an intriguing scenario where the flux rope may continue to rise through the repeated reconnections within the sheared arcade, ultimately reaching the height of the torus unstable region, thereby positioning the rope as a viable candidate for the eruption. It is important to note that the ascent of the flux rope is frequently linked to the appearance or cancellation of photospheric magnetic flux \citep{2011A&A...526A...2G,2014ApJ...789...46K,2019AGUFMSH33B3390T}, or to a destabilization effect caused by magnetic reconnection \citep{2016ApJ...830...80K, 2023A&A...677A..43P}. In this regard, we analyzed the temporal progression of the photospheric magnetic flux beginning 4 hours before the flare onset and continuing throughout the entire duration of the flare. Notably, we observed no distinct evidence of flux cancellation or emergence within the specified time period. This further supports the proposed scenario in which the evolution of the flux rope is driven by ongoing magnetic reconnection.    
In panel (c), we depict the $logQ$ in the $y$-constant plane, overlaid with the transverse field of the 3D null (derived by setting $B_y=0$ in 
$\textbf{B}$ for the yellow field lines), which envelopes the leg of the flux rope (refer to the inset of Figure \ref{initial_exp}(b)). It is important to highlight the elevated $logQ$ values in the vicinity of the neutral point \citep{2021SoPh..296...26K}. Consequently, reconnection within the 3D null is expected, which may facilitate the removal of the strapping field lines above the flux rope and promote its ascent \citep{Maurya2026}. }

\begin{figure}
    \centering
    \includegraphics[width=1.0\textwidth]{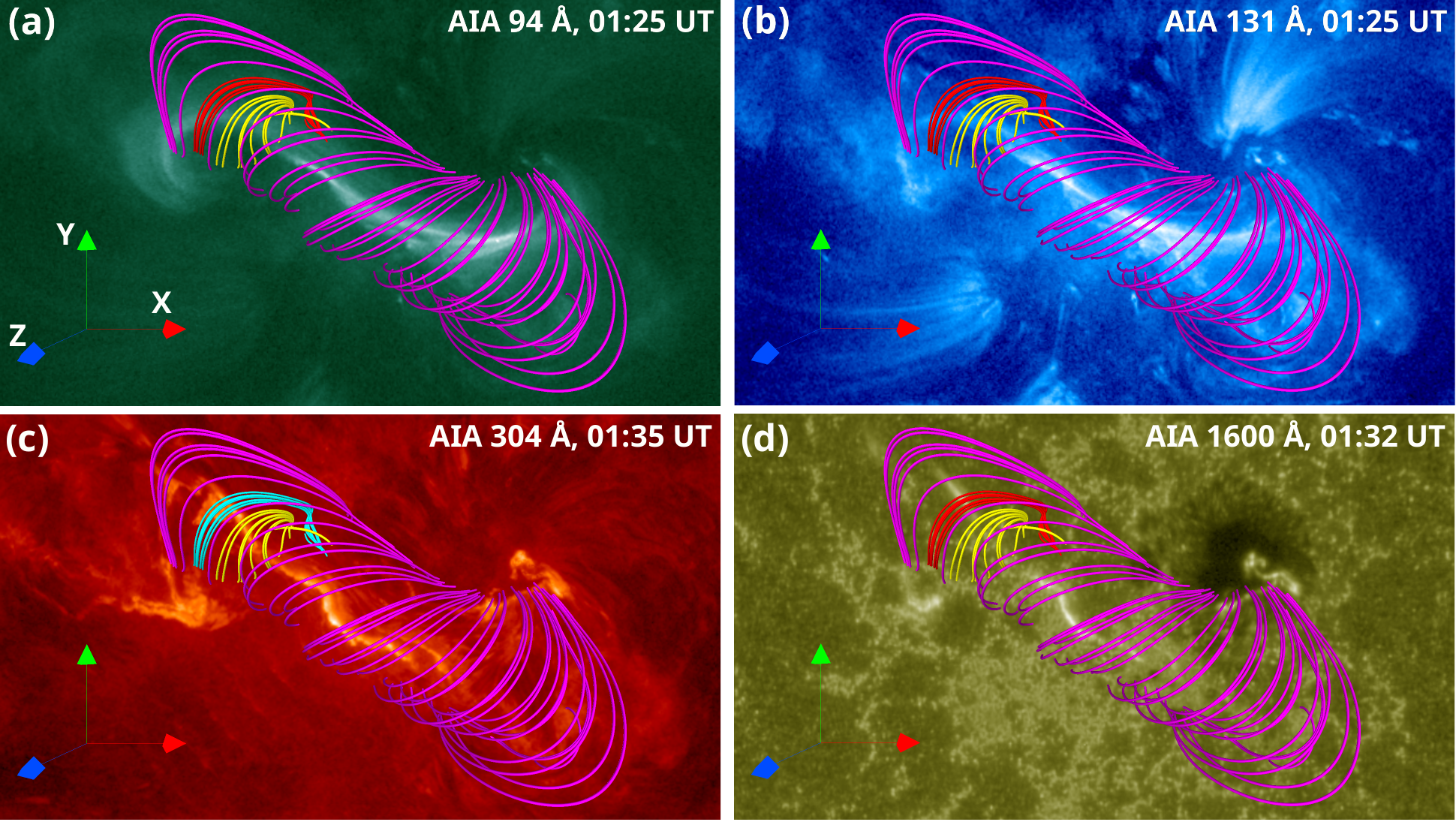}
    \caption{Panels (a) and (b) illustrate the extrapolated field lines overlaid with AIA 94{\AA} and 131 {\AA} images at the flare onset time at 01:25 UT. Panel (c) shows the field lines superimposed with 304 {\AA} image at 01:35 UT. To enhance visibility, the flux rope is depicted in cyan color in panel (c). Panel (d) depicts the extrapolated field lines on 1600 {\AA} image at 01:32 UT   }
    \label{exp-obs}
\end{figure}

Figure \ref{exp-obs} displays the extrapolated field lines superimposed with the AIA images. Here the background AIA images are chosen based on the clear appearance of the flare brightenings and  the ribbons to show the morphological correspondence with the extrapolated field lines. Panels (a) and (b) plot the field lines overlaid with the AIA 94 {\AA} and 131 {\AA} images at the time of flare onset. Significantly, the field lines of the sheared arcade (depicted in pink) closely follow the sigmoidal brightening (see Figures \ref{94_131_obs}(b) and (d)). It is crucial to emphasize that these field lines exhibit considerable shearing and twisting. As a result, they facilitate strong field-aligned currents, which may intensify over time due to the gradual movement of the foot-points of the field lines. The dissipation of these currents could lead to Joule heating of the plasma, potentially explaining the EUV emissions responsible for the formation of the sigmoidal brightening \citep{2020ApJ...903..129P}.

Furthermore, panels (c) and (d) of Figure \ref{exp-obs} superimpose the extrapolated field lines on the AIA 304 {\AA} and 1600 {\AA} images, respectively. From the panels, it appears that the reconnection inside the 3D null (in yellow) can cause the brightening BR1 (Figure \ref{304_obs}) and the ribbon R1 (Figure \ref{1600_obs}).  Moreover, the potential reconnection inside the sheared arcade beneath the flux rope is expected to accelerate the charged particles along its field lines that may contribute to the brightenings BR1 and BR2  (Figure \ref{304_obs}) and the flare ribbons R1 and R2 (Figure \ref{1600_obs}). Later, the onset of reconnection in QSL1 can contribute to the extension of BR2 (yellow arrow in Figure \ref{304_obs}(c)). The serpentine brightening BR3 and associated ribbon R3 can be attributed to the reconnection in the QSL2 and QSL4 (Figure \ref{initial_exp}(d)).  Furthermore the reconnection in the QSL3 can explain the appearance and extension of hook shaped brightening BR4 (Figure \ref{304_obs}(d)) and the ribbon R4 (Figure \ref{1600_obs}(b)-(c)). Moreover, the flux rope is located near the dark structure F observed in 304 {\AA} (see Figure \ref{304_obs} and Figure \ref{exp-obs}(c)) --- suggesting that the rope represents the magnetic configuration of the dark structure. It is crucial to emphasize that significant magnetic structures, including the flux rope and the 3D null, are situated in the vicinity of the ribbon R1. Consequently, it is anticipated that magnetic reconnections will occur more efficiently in this area, which may account for the substantial HXR emission originating from the area (Figure \ref{euv_uv_hxr_qpp}).

Interestingly, the proposed recurrence of magnetic reconnection at the 3D null as well as within the sheared arcade may play a crucial role in the observed QPPs during the impulsive phase of the flare \citep{2016ApJ...823..150T}.  The reconnection can potentially lead to the repetitive acceleration of non-thermal electrons, which may cause the QPPs. This suggested mechanism for the QPPs is consistent with a recent study by \citet{2025JGRA..13033772L}.

\section{Summary and Discussion}
\label{summ}
In this study, we investigate the onset of X1.8 class flare and associated QPPs in NOAA AR 13848 on 2024 October 9. For this purpose, we have taken the multi-wavelength observations from AIA, HMI/SDO, and HEL1OS/Aditya-L1 into account. In addition to this, we supplement these observations with a NFFF extrapolation prior to flare onset to identify the key magnetic features in the flaring region and their role in flare development.

Utilizing the multi-wavelength observations, we document the appearance of a sigmoidal brightening at the flare onset (Figure \ref{94_131_obs}). With time, four distinct flare brightenings BR1-BR4 (Figure \ref{304_obs}) and ribbons R1-R4 (Figure \ref{1600_obs}) with different morphology are found to develop in the flare kernel. Moreover, we observe the oscillation in the HXR light curve, which shows more than three distinct peaks during the impulsive phase of the flare (Figure \ref{QPP}). The Fourier analysis indicates that the oscillations can be categorized as QPPs, exhibiting a predominant period of approximately $4 \pm 0.6$ minutes. Likewise, the global wavelet analysis also records the existence of QPPs with a period of around $4 \pm 0.7$ minutes in the HXR emission. This duration exceeds the seconds to tens of seconds periodicities typically linked to rapid sausage-mode oscillations or plasmoid dynamics \citep{2010ApJ...723.1651K,2024ApJ...970...77L}, and it is similar to the several-minute timescales often connected with magnetic reconnection \citep{2009SSRv..149..119N,2019A&A...621A.106T,2024ApJ...973...74K,DongLi2025}. Given that HXR emission is intricately tied to particle acceleration and energy release, the observed near 4-minute periodicity may indicate a repetitive or episodic energy-release mechanism during the flare {\citep{Clarke_2021}}. Nevertheless, the period by itself does not definitively determine the underlying QPP mechanism, as similar timescales can result from various physical processes \citep{2023BAAS...55c.181I,2024ApJ...970...77L}.

To locate the QPP producing region, the temporal evolution of EUV and UV emissions from each of the brighten regions are compared with the HXR time series. Importantly, the pulsations in the temporal evolution of the ribbon R1 nearly coincident with the HXR pulsations (Figure \ref{euv_uv_hxr_qpp}(a)). Furthermore, the 20-50 keV HXR source contours are predominantly found to overlap with the R1 in AIA 1600 {\AA} (Figure \ref{euv_uv_hxr_qpp}(b)). This suggests that the QPPs detected in the HXR emission are primarily generated from the area of the ribbon R1.

The magnetic field configuration of the flaring region obtained from the NFFF extrapolation reveals the existence of key magnetic features such as a sheared arcade, a low-lying 3D null and a flux rope (Figure \ref{initial_exp}). Notably, the 3D null and the flux rope are located in the region of the ribbon R1. In addition, we identify high $logQ$ regions at the lower boundary of the flaring region, which are then categorized in four QSLs. The sigmoidal brightening is associated with the sheared arcade. 
The reconnection at the 3D null is expected to play a role in the appearance of brightening BR1/ribbon R1. In addition, notable is the presence of high $|\textbf{J}|/|\textbf{B}|$ inside the arcade below the flux rope (Figure \ref{recn}), which favors the onset of reconnection within the field lines of the arcade. The reconnection can potentially contribute to the brightenings BR1 and BR2 along with the associated ribbons R1 and R2 (Figure \ref{exp-obs}). It is notable that such reconnections when repeated over time, may contribute to the ascent of the flux rope and may lead to its eventual destabilization \citep{2016ApJ...830...80K}. Interestingly, the reconnection at the 3D null can further facilitate the potential eruption of the flux rope. Moreover, the expansion of the brightenings along with the appearances of additional flare brightenings BR3 and BR4 are suggested to be caused by the slipping reconnection in the QSLs.

The recurring reconnection episodes involving the 3D null point and the flux rope, which are found to be responsible for the formation of ribbon R1, may also contribute to the generation of the QPPs observed in the HXR waveband. The repeated occurrence of these reconnections can maintain a intermittent release of energy. This can cause a repeated energization and effective particle acceleration as proposed in previous works  \citep{2021ApJ...910..123C,2025JGRA..13033772L,2026NatAs..10...13L,2026NatAs..10...54A}.

Overall, this work proposes that the recurrent magnetic reconnection in the various magnetic features such as a sheared arcade, a 3D null and QSLs can play a central role in initiating the X-class flare and causing the QPPs. On the flip side, the present work is limited in identifying the exact physical origin of the observed QPP period directly from the extrapolated magnetic configuration. An MHD simulation demonstrating the time evolution of the magnetic field combined with observations of the erupting structure and reconnection rate is required to establish the detailed physical mechanism of the QPPs and associated periodicity. In future, we plan to advance this work to explore the energetics involved in the QPPs by incorporating such MHD simulations.

\begin{acks}
We acknowledge the use of the visualization software VAPOR 
(\url{www.vapor.ucar.edu}) for generating relevant graphics. The computations were performed on high performance computer of Department of Physics, Patna University, Patna, India. Data and images are courtesy of NASA/SDO and ISRO/Aditya L1 and the HMI, AIA and HEL1OS science teams. SDO/HMI as well as Aditya L1/HEL1OS are joint efforts of many teams and individuals to whom we are greatly indebted for providing the data. We also thank the reviewers
for providing insightful suggestions and comments that led to the significant improvement of
this paper.
\end{acks}

\begin{authorcontribution}
P.K., S.K, and S. analyzed the observations. P.K. and S.R.K. prepared the plots. P.K. and S.K. wrote the main draft of the paper. S.K. S. and Y.K.M contributed to the interpretation of the results. All authors did a careful proofreading of the text and improved the quality of the paper.
\end{authorcontribution}

\begin{fundinginformation}
S.K. and S. would like to acknowledge the support from the ANRF-SERB project No. SUR/2022/000569. S.K. also acknowledges support from Patna University project No. RDC/MRP/07.
\end{fundinginformation}

\begin{dataavailability}
The data that support the findings of this study are available from the
corresponding author upon reasonable request.
\end{dataavailability}

\begin{ethics}
\begin{conflict}
The authors declare that they have no conflicts of interest .
\end{conflict}
\end{ethics}

\bibliographystyle{spr-mp-sola}
\bibliography{ref}

\end{document}